\documentclass[12pt,amsmath,floats,floatfix,nofootinbib,secnumarabic, superscriptaddress]{revtex4}
\usepackage{graphicx} \usepackage{bm} \usepackage{epsfig}
\usepackage{pstricks}
\usepackage{setspace}
\usepackage{graphicx}
\usepackage{bm}
\usepackage{hyperref}
\usepackage{epsfig}
\usepackage{color}
\usepackage{colordvi}
\newcommand{\beq}{\begin{equation}}
\newcommand{\eeq}{\end{equation}}
\newcommand{\bea}{\begin{eqnarray}}
\newcommand{\eea}{\end{eqnarray}}
\newcommand{\bit}{\begin{itemize}}
\newcommand{\eit}{\end{itemize}}

\newcommand{\nn}{\nonumber}

\newcommand{\be}{\begin{equation}}
\newcommand{\ee}{\end{equation}}

\newcommand{\ber}{\begin{eqnarray}}
\newcommand{\eer}{\end{eqnarray}}
\newcommand{\beal}{\begin{align}}
\newcommand{\eal}{\end{align}}
\newcommand{\bes}{\begin{split}}
\newcommand{\es}{\end{split}}

\newcommand{\calo}{{\cal O}}

\begin{document}

\title{\Large The consistency condition for the three-point function\\ in dissipative single-clock inflation}

\author{Diana L\'opez Nacir}
\affiliation{Departamento de F\'isica, Facultad de Ciencias Exactas y Naturales, UBA and IFIBA, CONICET. Ciudad Universitaria, Pabell\'on 1, 1428, Buenos Aires, Argentina.\vskip 0.25cm}
\author{Rafael A. Porto}
\affiliation{School of Natural Sciences, Institute for Advanced Study, Einstein Drive, Princeton, NJ 08540, USA\vskip 0.25cm}
\affiliation{Department of Physics \& ISCAP, Columbia University, New York, NY 10027, USA \vskip 0.25cm}
\author{Matias Zaldarriaga}
\affiliation{School of Natural Sciences, Institute for Advanced Study, Einstein Drive, Princeton, NJ 08540, USA\vskip 0.25cm}

\begin{abstract}
We generalize the consistency condition for the three-point function in single field inflation to the case of dissipative, multi-field, {\it single-clock} models. 
We use the recently introduced extension of the effective field theory of inflation that accounts for dissipative effects, to provide an explicit proof to leading (non-trivial) 
order in the generalized slow roll parameters and mixing with gravity scales. Our results  illustrate the conditions necessary for the validity of the consistency relation in situations with many   degrees of freedom relevant during inflation, namely that there is a preferred clock. Departures from this condition in forthcoming experiments would rule out not only single field but also a large class of multi-field models. 

\end{abstract}

\maketitle

\newpage
\tableofcontents
\newpage

\section{Introduction}

The {\it squeezed limit} of the three-point function of curvature perturbations, geometrically:
\beq
\langle \zeta_{k_L}\zeta_{k_S}\zeta_{k_S}\rangle_{k_L \to 0} \leftrightarrow \parbox{58mm}{\includegraphics{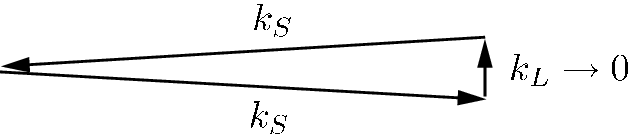}} \nonumber ~~~~~,
\eeq
is tailor made to probe the dynamics of the early universe, due to its intrinsic dependence on the very mechanism that produces such perturbations. It is observationally relevant in many contexts, for example in measurements of large scale structure \cite{sqz1,sqz2} or CMB $\mu$-distortion \cite{mud}. For models with only one {\it light} dynamical field that is relevant during inflation, there is a consistency condition which relates the three-point function in the squeezed limit to  deviations from scale invariance in the power spectrum  \cite{maldacena,consist1,consist2,cremi,consist3}. This relation can be written more precisely as
\beq\label{consist}
\lim_{k_1\to 0} \langle\zeta_{k_1}\zeta_{k_2}\zeta_{k_3}\rangle=-(2\pi)^3\delta^3\left(\sum_i {\bf k}_i\right)P_{\zeta}(k_L)P_{\zeta}(k_S)\left[(n_s-1) + O\left(k_L^2/ k_S^2\right)\right],\eeq
where ${\bf k}_1\to {\bf k}_L$ and $({\bf k}_2-{\bf k}_3)/2 \to {\bf k}_S$, for the long- and short-wavelength modes respectively. Also $P_{\zeta}(k_i)$ is the power spectrum, i.e. $\langle \zeta_{k_i}\zeta_{k_j}\rangle = \delta^3({\bf k}_i+{\bf k}_j) P_\zeta(k_i)$, and $n_s-1$  is the tilt 
\beq
n_s-1=\frac{d\log{k^3P_{\zeta}(k)}}{d\log k},
\eeq
such that $\langle \zeta \zeta \rangle \simeq k^{-3+n_s-1}$. Alternatively, defining $F(k_1,k_2,k_3)$ by
\be \langle\zeta_{k_1}\zeta_{k_2}\zeta_{k_3}\rangle=(2\pi)^3\delta^3(\sum_i{\bf k}_i) F(k_1,k_2,k_3),\ee
and using the  definition of  the parameter $f^{\rm sq}_{\rm NL}$
\begin{equation}\label{squ}
f_{\rm NL}^{\rm sq} \equiv \lim_{k_3\to 0}\frac{5}{6} \frac{F(k_1,k_2,k_3)}{(P_{\zeta}(k_1)P_{\zeta}(k_2)+P_{\zeta}(k_1)P_{\zeta}(k_3)+P_{\zeta}(k_2)P_{\zeta}(k_3))},
\end{equation}  
the consistency relation in the exact squeezed limit can be written as
\be
\label{fnlsq}
f_{\rm NL}^{\rm sq}=-\frac{5}{12}(n_s-1).
\ee

The expression in Eq. (\ref{consist}) follows in the long-wavelength limit, $k_L \to 0$, as an expansion in powers of $k_L/k_S$ which is valid to all orders in the slow roll parameters \cite{consist1,consist2,cremi,consist3}. Therefore, Eq. (\ref{consist}) would remain valid regardless of the value for $n_s-1$, even for hypothetical scenarios with large departures from scale invariance. 

Eq. (\ref{consist}) is nothing but the manifestation of the fact that the long mode has no dynamical effect on the short modes, and that the three-point function in the squeezed limit is in a sense just an artifact of the coordinates being used. If rather than co-moving coordinates we perform a transformation to a more {\it physical}  set of coordinates, i.e. $x^i_{\rm phys} = e^{\zeta_L} x^i$, or $k^S_{\rm phys} = e^{-\zeta_L} k_S$ then the amplitude of the small scale power would become independent of the long mode. 
There is no three-point function in these coordinates. In gravity physical effects are encoded in the curvature and thus one expects that the first correction in Eq. (\ref{consist}) scales as $(k_L/k_S)^2$.  More detailed versions of this argument can be found in  \cite{consist1,consist2,consist3,cremi}. 


The validity of the consistency relation is evident in cases where heavy (or decoupled) fields are present, but becomes non-trivial in situations where light excitations are produced during inflation which subsequently decouple, or in general when additional degrees of freedom (ADOF) produce negligible direct contributions to $\zeta$, but significantly alter the dynamics of the (single) clock which controls the end of inflation. 

Scenarios where ADOF couple to the clock, but do not contribute to density perturbations, are tantamount to study dissipative effects during inflation, such as warm \cite{warm} and trapped \cite{trapped} inflation, which were recently incorporated in \cite{diseft} (based on ideas originated in \cite{dis1,dis2}) within the effective field theory (EFT) of inflation formalism developed in 
\cite{eft1,Creminelli:2006xe,Senatore:2009gt, Senatore:2009cf, Senatore:2010jy, Bartolo:2010bj, Bartolo:2010di,dan2,strongc,dan2new}. 

An explicit proof of the consistency condition for single field inflation, valid at leading non-trivial order in the slow roll parameters, was given in \cite{consist2} using the EFT approach \cite{eft1}. Our purpose in this paper is to extend the results in \cite{consist2} to single-clock models with dissipative ADOF. Proving the consistency condition in these type of scenarios is more challenging than in single field inflation due to the presence of ADOF whose detailed dynamics, and couplings, remain vastly unknown. However, thanks to the EFT approach \cite{eft1,diseft}, we will show that under certain circumstances the consistency relation holds to leading (non-trivial) order in the generalized slow roll parameters and mixing with gravity scales. Departures from this condition in forthcoming experiments would rule out not only single field, but also a large class of multi-field single-clock models.\\

This paper is organized as follows. In the next section we review the EFT formalism introduced in \cite{eft1,diseft}, in particular: The assumptions we make about the properties of the noise and response for the ADOF; the constraint equations in the presence of ADOF and the constancy of $\zeta$ outside the horizon; the computation of the power spectrum and tilt. Then, in sec. \ref{sqz} we explicitly show the validity of the consistency relation to leading (non-trivial) order in the generalized slow roll parameters, first in the limit $M_p\to \infty$ and later on including the mixing with gravity. As expected, in sec. \ref{genr} we show that even in the presence of ADOF a long-wavelength mode amounts to a rescaling of the coordinates, as in single field inflation. 

We concentrate in models with scalar ADOF, although we discuss towards the end in sec. \ref{other} how to generalize our analysis for vector and tensor operators. We relegate details and more technical points to appendices, including the different contributions from each source term and $(k_L/k_S)^2$ scaling in the three-point function. Everywhere we set $c=\hbar=1$ and adopt the mostly plus sign convention.

\section{Effective Field Theory set up}

Here we introduce the necessary elements of the EFT formalism developed in \cite{diseft} to study dissipative effects during inflation.\footnote{The EFT of multi-field inflation with many light degrees of freedom was studied in \cite{multieft}.} We concentrate on the features which are relevant to analyze the squeezed limit of the three-point function.

\subsection{The effective action}

In \cite{diseft} dissipative effects are incorporated via the coupling to a set of (scalar, vector, tensor) composite operators $\calo^{\mu\nu\ldots}$ in the effective action, which is constrained solely by invariance under all the symmetries of the long distance physics.

The so-called {\it unitary gauge} is defined such that the perturbations of the preferred clock vanish. In this gauge the unit vector perpendicular to surfaces of constant time  ${t}$ takes the form  $ n_{\mu}=-\delta^{0}_{\mu} (-g^{00})^{-1/2}$, and the  extrinsic curvature of the surfaces is $K^{\mu}_{\nu}=\hat{g}^{\mu\rho}\nabla_{\rho}n_{\nu}$, where  $\hat{g}_{\mu\rho}=g_{\mu\rho}+n_{\mu}n_{\rho}$ is the induced spatial metric. In addition, one can choose coordinates so that the metric for fluctuations around the quasi-deSitter background is given by
\be ds^2=-N^2dt^2+ a^2(t)\delta_{ij} e^{2\zeta} (dx^i+N^i dt) (dx^j+N^j dt).\ee
In this paper we ignore tensor perturbations.\\

The action in the unitary gauge reads \cite{diseft,eft1}
\begin{eqnarray}
\label{act1}
S&=&\frac{M_p^2}{2}\int d^4 x\sqrt{-g} R+\frac{1}{2}\int d^4 x \sqrt{-g}(\overline{p}-\overline{\rho}-(\overline{p}+\overline{\rho}) g^{00})\nonumber\\
&+&\frac{1}{2} \int d^4 x \sqrt{-g}\,M_2^4(t) (1+ g^{00})^2-\frac{1}{2}\int d^4 x \sqrt{-g}\,\overline{M}_1^3(t)\delta K_{\mu}^{\mu}(1+ g^{00})\nonumber\\
&-&\int d^4 x \sqrt{-g}\, f(t)\calo +S_{\calo}+\cdots,\label{action}
\end{eqnarray} where $M_{p}^2=(8\pi G_N)^{-1}$, and $\calo$ is a scalar composite operator associated with the dissipative degrees of freedom and $S_{\calo}$ represents the action for this sector, which we do not need to specify. The dots stand for higher derivative corrections. Also, bared quantities denote their unperturbed value. Throughout the paper we assume the presence of an approximate shift symmetry, such that functions of time appearing in the action change very little in a Hubble time. This will be the basis of our generalized slow roll approximation. Since we are only interested in the squeezed limit of the three-point function the higher derivative terms are irrelevant.\\

The field $\pi$, which characterizes the perturbations of the preferred clock, is then introduced via the St\"uckelberg's trick, i.e. $t \to  t+\pi$ \cite{eft1}. It is easy to show that the relation between $\pi$ and $\zeta$ is the same as in single field inflation. Taking into account that in the $\pi$-gauge the metric takes the form
\be ds^2=-N^2dt^2+a^2(t)\delta_{ij} (dx^i+N^i dt) (dx^j+N^j dt),\ee
the relationship becomes, up to second order in the perturbations \cite{maldacena},
\begin{eqnarray} 
\zeta&=&-H\pi+H\pi\dot{\pi}+\frac{1}{2}\dot{H}\pi^2+\alpha,\label{relationzetapi}\\
\alpha&=&\frac{1}{a^2}\left(-\partial_i\pi\partial_i\pi+\partial^{-2}\partial_i\partial_j(\partial_i\pi\partial_j\pi)\right).
\end{eqnarray} 

Notice we restricted ourselves to scalar couplings for the ADOF of the form $f(t)\calo$, which naturally produce a $\gamma\dot\pi$ dissipative term when the Green's function obeys ${\rm Im}G^{\calo}(\omega)\sim \omega$ \cite{diseft}. There are, however, other type of scalar couplings, such as $\tilde\calo g^{00}$. Even though these produce terms involving derivatives of $\pi$, they can also contribute at leading order in the slow parameters in cases when ${\rm Im}G^{\tilde\calo}(\omega)\sim 1/\omega$, as it is required for $\gamma\dot\pi$ dissipation \cite{diseft}. We will return to this possibility, as well as having vector and tensor couplings, later on in sec. \ref{other}.\\

\subsection{Noise \& Response}\label{noiseresp}

Notice that, neglecting metric perturbations, the presence of the ADOF affects the clock only through the coupling between $\pi$ and $\calo$. As explained in \cite{diseft}, this introduces dissipation and noise effects into the dynamics of $\pi$ which can be characterized by the splitting of $\calo$ into pieces:\footnote{We use $\delta\calo_{\cal S}$, instead of $\delta\calo_S$, in order to avoid confusion with ${}_S$ being used to describe short modes.}
\beq
\label{calotxsr}
\calo (t,{\bf x})= \bar\calo(t)+ \delta\calo_{\cal S}(t,{\bf x}) + \delta\calo_R (t,{\bf x}),
\eeq where $\bar\calo(t)$ is the background value, $\delta\calo_R (t,{\bf x})$ represents the change in the expectation value that results as a response to the interaction with $\pi$, and $\delta\calo_{\cal S}(t,{\bf x})$ corresponds to a stochastic noise. 
We will work in the approximation where both the response part $\delta\calo_R(t,{\bf x})$ and the two-point correlator of the noise $\langle\delta\calo_{\cal S}(t,{\bf x})\delta\calo_{\cal S}(t',{\bf x}')\rangle$
are local in time (namely we neglect any memory effect), and we assume an expansion in spatial derivatives holds. Then, in unitary gauge, we have for the noise 
\beq
\label{twop1}
\langle \delta\calo_{\cal S}(t,{\bf k})\delta\calo_{\cal S}(t',{\bf q})\rangle \simeq \frac{\nu_{\calo}(t)}{\sqrt{-g}}\delta(t-t') (2\pi)^3 \delta^{(3)}({\bf q+\bf k}).
\eeq
On the other hand, for the response part we expect in general a (local) relationship of the sort \cite{diseft}
\beq
\label{respo0}
\bar\calo+ \delta \calo^u_R \simeq  \Lambda_\calo(t)~F\left[\sqrt{-g^{00}}, K^\mu_\mu, t \right],
\eeq
where $F$ is some generic function and $\Lambda_\calo$ provides the necessary scale such that \beq \bar\calo(t)~=~\Lambda_\calo(t) F[1, 3H, t].\eeq

Because of the time dependence of the background, it is convenient to define the operator \cite{diseft}
\beq
\label{calou}
\delta\calo^s_R(\pi) \equiv \left(\calo^\pi_R- \bar{\calo}(t+\pi)\right),
\eeq
so that going from the unitary to the $\pi$-gauge we have $\delta \calo_R^u \to \delta \calo^s_R(\pi)$. ($\calo$ is a scalar operator which does not introduce any $\pi$'s. This replacement is then required to cancel out the expansion in $t+\pi$ from the background value $\bar\calo(t)$.) 

Using the expression in Eq. (\ref{respo0}) we have
\beq
\label{respo1}
\delta\calo^s_R (\pi) = \Lambda_\calo(t+\pi)F\left(n^\mu \partial_\mu(t+\pi), K^\mu_\mu(t+\pi),t+\pi \right)-\bar\calo(t+\pi)
\eeq
where 
\beq \label{nmu} n^{\mu}=\frac{- g^{\mu\nu}\partial_{\nu}{(t+\pi)}}{\sqrt{-g^{\nu\rho}\partial_{\nu}{(t+\pi)}\partial_{\rho}{(t+\pi)}}}.
\eeq

Using analyticity for small perturbation around the background $\bar n^\mu = \delta^{\mu0}$, Eq. (\ref{respo1}) can be expanded as\footnote{For example let us consider a function 
$\calo_R=F(\Lambda_1 \sqrt{X})$, where $X \equiv -g^{\mu\nu} \partial_\mu(t+\pi)\partial_\nu (t+\pi)$. Then, $\bar X=1$ and 
$F(\Lambda_1 (t+\pi) \sqrt{1 + \delta X}) \simeq \bar\calo(t+\pi) + \frac{F'(\Lambda_1)}{2}\Lambda_1(t+\pi) \delta X + \cdots$, with $\delta X = -(1+g^{\mu\nu} \partial_\mu(t+\pi)\partial_\nu (t+\pi))$. Hence we obtain Eq. (\ref{responseful}) for the expansion of $\delta\calo_R^s$.}
\bea\label{responseful}
\delta\calo_R^s (\pi) &=&\left[\frac{-V_\calo(t+\pi)}{2}(1+g^{\mu\nu}\partial_{\mu}(t+\pi)\partial_{\nu}(t+\pi))+\right.\\ && \left. 
 W_\calo(t+\pi)(1+g^{\mu\nu}\partial_{\mu}(t+\pi)\partial_{\nu}(t+\pi))^2+\cdots\right].\nonumber
\eea

The ellipses in Eq. (\ref{responseful}) account also for terms proportional to the extrinsic curvature which one can show at most contribute to a renormalization of the speed of sound, and therefore does not alter the proof of the consistency condition. This expression encodes all the information about the coupling between $\calo$ and $\pi$, including all possible non-linear interactions.\footnote{In other words, Eq. (\ref{responseful}) includes both: non-linear response applied to linear forces, as well as linear response applied to non-linear forces in $\pi$.} Notice that the RHS of Eq. (\ref{responseful}) is (approximately) invariant under $\pi \to \pi+c$ provided $V_\calo(t),W_\calo(t),\ldots$ are slowly varying functions of time. From now on we will assume that is the case for all time dependent functions in the problem, including $\bar\calo(t)$. This is what we refer as the existence of an approximate (emergent) shift symmetry in $\pi$ \cite{diseft}.

Note also that we write the metric components in the contractions explicitly. This is because, as we will discuss momentarily, they play a crucial role to guarantee the constancy of $\zeta$ outside the horizon, as well as to incorporate the mixing with gravity.\\

In what follows we will assume an expansion as in Eq. (\ref{responseful}) applies, not only to $\calo$, but all operators that derive from the ADOF, e.g. $(T^0_0)_{\calo}$ etc. 

\subsection{Constraint equations \& Constancy of $\zeta$ outside the horizon}\label{constr}

Using the expression in Eq. (\ref{act1}) the background equations are given by: 
\begin{eqnarray}
&& 3H^2 M_p^2=\bar{\rho}+\bar{\rho}_{\calo}+f(t)\bar\calo, \label{beqmet}\\
&&\dot{\bar\rho}+3H(\bar\rho+\bar p)+\dot{f}\bar{\calo}=0,\label{beqpi}\\
&&\dot{\bar\rho}_{\calo}+3H(\bar\rho_{\calo}+\bar p_{\calo})+f\dot{\bar{\calo}}=0,\label{bacO}
\end{eqnarray} where the last one is obtained by combining the second  and the time derivative of the first,
 and we used that the background stress tensor for the ADOF obtained from $S_{\calo}$ takes the perfect fluid form
\beq
{\bar T}_{\mu\nu}^{\calo} = (\bar \rho_{\calo}+\bar p_{\calo}){\bar n}_\mu {\bar n}_\nu + \bar g_{\mu\nu} \bar p_{\calo}.
\eeq 

To linear order in the perturbations the constraint equations become  
\begin{eqnarray}
&&2H M_p^2\partial_i\delta N+ \frac{M_p^2}{a^2}\left(\partial_i\partial_jN_i -\partial_i\partial_iN_j\right)-(\bar p+\bar\rho)\partial_i\pi-\overline{M}_1^3\partial_i(\delta N-\dot{\pi})+  \delta B^{(1)}_i=0,\\
&&-6 H^2\delta N M_p^2-2 H \partial_iN^iM_p^2+\overline{M}_1^3\left( \partial_iN^i+3H(\delta N-\epsilon H\pi)+\frac{\triangle\pi}{a^2}\right)\nonumber\\
&&+(N_c+3H\overline{M}_1^3)(\delta N-\dot{\pi})-\dot{\bar{\rho}}\pi-\dot{f}\bar{\calo}\pi-f\delta\calo^{(1)}+ \delta A^{(1)}=0, 
\end{eqnarray}  where the superscript ${}^{(n)}$ indicates the order in the perturbations, and
$\delta A^{(1)}$ and $\delta B^{(1)}_i$ are the linearized part of
\begin{eqnarray}
A&=&\frac{1}{\sqrt{h}}\frac{\delta S_{\calo}}{\delta N}, 
\\
B_i&=&\frac{1}{\sqrt{h}}\frac{\delta S_{\calo}}{\delta N^i}.
\end{eqnarray} Note that for the background we have $\bar{A}=-\bar \rho_{\calo}$ and $\bar{B}_i=0$. We also introduced 
\beq 
N_c \equiv (\overline{p}+\overline{\rho}+4M_2^4) \label{eqnc},
\eeq 
which will play the role of our normalization scale for $\pi$ later on.\\

Solving for the lapse and shift we find (for $H^2M_p^2\gg N_c,H\overline{M}_1^3$)
\begin{eqnarray}
\delta N&\simeq&\frac{\bar p+\bar\rho}{2 H M_p^2}\pi- \frac{\delta B^{(1)}}{2HM_p^2}-\frac{\overline{M}_1^3(\dot{\pi}-\epsilon H\pi)}{2HM_p^2},\\
\partial_iN^i&\simeq&-\frac{1}{2 H M_p^2}\left( N_c(\dot{\pi}-\epsilon H\pi)-3 H \delta B^{(1)}-\overline{M}_1^3\frac{\triangle\pi}{a^2}+f\delta \calo^{(1)}- \delta A^{(1)}\right),
\end{eqnarray}
where $\delta B^{(1)}=\partial^{-2}\partial_i \delta B^{(1)}_i$ and we used Eq. (\ref{beqpi}).\\

It is now convenient to rewrite these equations in terms of $\delta\calo^s$, as in Eq. (\ref{calou}), and accordingly:
\begin{eqnarray}
\delta A^s&\equiv&\delta A+{\dot{\bar\rho}}_{\calo}{\pi},\\
\delta B^s &\equiv &\delta B+(\bar{p}_{\calo}+\bar{\rho}_{\calo})\pi.
\end{eqnarray}  The reason will be soon clear, but notice that at linear order these operators correspond to the (scalar) perturbations of $\calo$, $\left(T_0^0\right)_\calo$ and $\partial^{-2}\partial_i \left(T^0_i\right)_\calo$, respectively.\footnote{The reader may be more familiar with gauge transformations of the fluid variables associated to the linearized stress tensor of a fluid, 
for which  $T^0_0=-\delta\rho$,  $\partial^{-2}\partial_i T^0_i=(\bar{\rho}+\bar{p})\delta u$.}
Using these definitions, and Eq. (\ref{bacO}), we can then write
\begin{eqnarray}
\delta N&\simeq&\epsilon H \pi+ \frac{\delta B^s_{(1)}}{2HM_p^2}-\frac{\overline{M}_1^3(\dot{\pi}-\epsilon H\pi)}{2HM_p^2},\label{deltaN}\\
\partial_iN^i&\simeq&-\frac{1}{2 H M_p^2}\left(N_c(\dot{\pi}-\epsilon H\pi)-3 H \delta B^s_{(1)}-\overline{M}_1^3\frac{\triangle\pi}{a^2}+f\delta \calo^s_{(1)}- \delta A^s_{(1)}\right)\label{deltaNi}.
\end{eqnarray}

The expression in Eq. (\ref{deltaN}) closely resembles the relationship in single field inflation, except for the appearance of new terms proportional to perturbations of the ADOF. However, as we argue in the previous section, 
the response for this term must obey a local expansion similar to the one in Eq. (\ref{responseful}).\footnote{Here we ignore the extrinsic curvature contribution since 
$\delta K^\mu_\mu \simeq \epsilon c_s^2 \dot \zeta + O(\dot\zeta^2)$, and therefore it does not alter our results to the order we work here.} In other words, including the mixing with gravity to leading (non-trivial) order,
\beq
\label{deltbs1}
\delta B^s_{(1)} \simeq V_B(t) (\dot \pi - \delta N)  + \cdots ,
\eeq
such that ($M^4_B = H V_B$)
\beq 
\delta N\simeq \epsilon H \pi+ \frac{M_B^4}{2H^2M_p^2}(\dot\pi - \delta N)-\frac{\overline{M}_1^3(\dot{\pi}-\epsilon H\pi)}{2HM_p^2}.\label{deltaN2}
\eeq
Hence,  
\beq 
(\delta N-\epsilon H \pi)\simeq \frac{\overline\epsilon}{1+\epsilon_B} (\dot{\pi}-\epsilon H\pi) + \cdots,\label{deltaN3}
\eeq

where we defined $\epsilon_B \equiv \frac{M_B^4}{2M_p^2 H^2}$, $\epsilon_{\overline{M}_1}=\frac{H\overline{M}_1^3}{2H^2M_p^2}$, so that $\overline\epsilon=\epsilon_B-\epsilon_{\overline{M}_1}.$\\

The expression in Eq. (\ref{deltaN3}) now mimics what occurs in single field inflation. In that case the second term can be shown to vanish for long-wavelength modes since $\dot \zeta_L \simeq 0 \to \dot\pi_L \simeq \epsilon H\pi_L$. In order to use a similar reasoning we need to show that $\frac{d}{dt}(H\pi)\to 0$ in the presence of the ADOF. Or, in other words, that $\zeta$ is still conserved outside the horizon. But using Eq. (\ref{responseful}) we have (including the mixing term)
\beq  
\label{resps}
\delta\calo^s_R \simeq V_\calo(t)(\dot\pi-\delta N)+\cdots \simeq V_\calo(t) \left(1+\frac{\overline\epsilon}{1+\epsilon_B}\right) (\dot \pi - \epsilon H \pi) + \cdots ,
\eeq
which is proportional to $\dot \zeta$, thus it turns off when $\zeta$ is constant. This implies a constant $\zeta=-H\pi$ remains a solution. Therefore \beq \dot\pi_L \simeq \epsilon H \pi_L \eeq in the long-wavelength limit (see appendix \ref{constantz} for more details). Moreover, from here we conclude \beq \delta N_L \simeq \epsilon H \pi_L \label{dneps}.\eeq 
A similar argument can be used for Eq. (\ref{deltaNi}).\\

Note that in principle not only the response, but also the noise will enter in the above expressions for the lapse and shift. However $\langle \delta B_{\cal S}\rangle =0$, so that the noise part only affects $\pi$, and $\zeta$, as a stochastic source and it will not alter its constancy outside the horizon. This is because, within the local approximation and for slowly varying functions of time, the computation of the power spectrum depends on integrals of Green's functions which are solutions of the homogeneous equation away from the $\delta$-source. As long as this equation admits $\zeta= \rm constant$ as a solution, under our assumptions the noise will not modify such behavior.\\

Notice that in the previous analysis we did not require $\epsilon$ to be small, then the constancy of $\zeta$ also applies to FRW cosmologies. However, since we assume the ADOF do not modify significantly the background dynamics, we do require $\epsilon_B \ll 1$. 
More generally, by assumption, the response parts for the ADOF are functions of the geometry, namely $g^{00}, K^\mu_\mu$, and these tend to their unperturbed value when $\epsilon_B \to 0$. Therefore --in the case of stable pertubations in the absence of ADOF-- this property will not be modified by adding a small correction to the background equations.

One may wonder about the case when $\epsilon_B \simeq 1$. In that scenario the coupling to gravity in the $\calo$-sector cannot be treated perturbatively. However, following similar steps as above, one can still show that a constant $\zeta$ will stay a solution, although we cannot guarantee it will be the dominant one \cite{strongc}. Note this must be the case because $\zeta \to \zeta+ \lambda$ is a symmetry in an FRW background at linear order, since $\lambda$ can be absorbed into a re-scaling of the coordinates. Moreover, we expect this to remain valid under the assumption of local dynamics; since the response $\delta B_R^s$ will indeed turn off when $\dot\zeta=0$ in the $k_L \to 0$ limit, and there will be no contribution to the long-wavelength curvature perturbations from earlier times when the modes were inside the horizon. 

One could still ask if $\epsilon_B \simeq 1$ is even feasible during inflation, namely having an expanding background together with a scale invariant spectrum with constant curvature perturbations outside the horizon.\footnote{Note that in this case, however, the explicit proof of the consistency condition becomes more involved, since one cannot ignore the contribution from the ADOF to $\delta N$ for short wavelength modes (see sec. \ref{mix}). } 
First of all, the slow roll condition $(-\dot H/H^2) \ll 1$ might be violated by contributions from the energy density in the $\calo$-sector. We could conceivable devise a scenario where violations of slow roll occur during a short period of time. But that is not the spirit of most dissipative models, where the energy scales associated with the ADOF are smaller than $M_p^2\dot H$. We will not discuss modifications of this set up in this paper, and from now on assume $\epsilon_B \ll 1$.

\subsection{Power spectrum \& tilt}\label{stilt}

We now move on to the computation of the power spectrum, putting emphasis on all possible contributions to the tilt. Using Eq. (\ref{act1}) and expanding to quadratic order, the action for $\pi$ can be written as (ignoring higher derivative corrections)
\begin{eqnarray}\label{l2pi}
S_{\pi}=\int d^4x a^3\frac{N_c}{2}\left\{\dot{\pi}^2-c_s^2\frac{(\partial_i\pi)^2}{a^2}-\frac{\dot{f}}{N_c}\pi \left(\delta\calo^s_R + \delta\calo_{\cal S}\right)+\cdots\right\},
\end{eqnarray}
where
 \beq \label{ncs} c_s^2=\frac{(\overline{p}+\overline{\rho}+H\overline{M}_1^3)}{(\overline{p}+\overline{\rho}+4M_2^4)},
 \eeq
and $N_c$ was introduced in Eq. (\ref{eqnc}), which can be also written as $N_c=(\overline{p}+\overline{\rho}+H\overline{M}_1^3)/c_s^2.$\\
 
 As we mentioned, we assume the presence of an approximate shift symmetry such that functions of time appearing in the action change very little in a Hubble time 
(e.g. ${\ddot f \over \dot f H} \ll 1, \frac{\dot N_c}{H N_c}\ll 1$, etc), and also ignored terms proportional to the generalized slow roll parameters which do not alter the computation of the power spectrum to the order we work here. (We can then replace $\delta\calo_R^s(\pi)$ by $\delta \calo_R^\pi(\pi)$, since $\dot{\bar\calo}$ is suppressed by a slow roll parameter. We omit the superscripts below.)\\

Recall also that, due to the presence of ADOF, our field $\pi$ differs from the choice made in \cite{eft1}, denoted as $\tilde \pi$ in \cite{diseft}. The quadratic Lagrangian for $\tilde \pi$ is uniquely determined by the quasi-deSitter geometry, e.g. $H,\dot H$, whereas in Eq. (\ref{l2pi}) the normalization scale $N_c$ is a free parameter. (One can show $c_s^2 N_c \leq -2 M_p^2\dot H$ -- assuming the contribution from $\overline{M}_1$ is subleading -- if the stress energy tensor of the ADOF obeys the null energy condition \cite{diseft}.) Our choice of unitary gauge, however, guarantees the relationship between $\zeta$ and $\pi$ remains given by Eq. (\ref{relationzetapi}). Note that  this relation is altered if written in terms of $\tilde\pi$ since (schematically) $\tilde \pi \sim \pi +\delta \calo$. See the discussion in \cite{diseft} for more details.\\

The equation for $\pi$ that follows from Eq. (\ref{l2pi}) becomes
\be\label{piaLO0}
\ddot{\pi}_k+3H\dot{\pi}_k+\frac{k^2c_s^2}{a^2}\pi_k+\frac{\dot f}{N_c}\delta\calo_R^{(1)}(t,{\bf k})=-\frac{\dot f}{N_c}\delta \calo_{\cal S}(t,{\bf k}),
\ee  
where $\delta\calo^{(1)}_R$ is the response part at leading order in $\pi$ which, using Eq. (\ref{responseful}),
can be written as 
\beq
\label{ddff}
\delta\calo^{(1)}_R\simeq V_\calo (t)\dot\pi,~~{\rm with}~~V_\calo(t) \equiv \gamma N_c/\dot{f},\eeq 
such that 
\be\label{piaLO}
\ddot{\pi}_k+(3H+\gamma)\dot{\pi}_k+\frac{k^2c_s^2}{a^2}\pi_k=-\frac{\dot f}{N_c}\delta \calo_{\cal S}(t,{\bf k}).
\ee

The expression in Eq. (\ref{ddff}) makes manifest the presence of an approximate (emergent) shift symmetry at the level of the response so that, to leading order, we can neglect terms which do not involve derivatives. To compute the power spectrum we also need the two-point function for the noise which is given in Eq. (\ref{twop1}).

The solution of Eq. (\ref{piaLO}) to linear order is given by (neglecting the homogenous solution which dies off in the $\gamma \gg H$ limit) \cite{diseft}
\beq \label{pi1}\pi_1(\eta,{\bf k})=\frac{ \dot f k c_s}{N_cH^2}\int^{\eta}_{\eta_0} d\eta' G_{\gamma}(k c_s|\eta|,k c_s|\eta'|)\frac{\delta{ \calo}_S}{(k c_s{\eta'})^2 },\ee
where $\eta_0$ is an early enough initial time, and
\beq
\label{greengamma}
G_{\gamma}(z,z')=\frac{\pi}{2} z\left(\frac{z}{z'}\right)^{\nu-1} \left[Y_\nu(z) J_\nu (z')-J_\nu (z) Y_\nu(z')\right],
\eeq 
with $\nu=\frac{3}{2}+\frac{\gamma}{2 H}$, $z=-k c_s\eta$, and  $z'=-k c_s\eta'$. Then, from Eq.~(\ref{twop1}) we obtain
\beq \nonumber  P_{\pi}(k)\equiv \langle\pi_1(\eta,{\bf k})\pi_1(\eta,{\bf k})\rangle=\frac{\nu_{\calo}{\dot f}^2 }{N_c^2(kc_s)^3}\int^{z_0}_z dz' (G_{\gamma}(z,z'))^2.
\eeq

For $kc_s\eta\to 0$ and $k c_s \eta_0 \to -\infty$, and using $P_{\zeta}=H^2P_{\pi}$,  we find
\beq\label{pow1gammazeta}\Delta_{\zeta} \equiv k^3P_{\zeta}(k)=\frac{\nu_{\calo}^\star{\dot f}^2_\star H^2_\star }{{N_c^\star}^2{c_s^\star}^3}\frac{16^{\frac{\gamma_\star}{H_\star} } (\frac{\gamma_\star}{H_\star} +1)^3
 \Gamma \left(\frac{\gamma_\star +H_\star}{2H_\star}\right)^4}{\pi  \Gamma ( \frac{2\gamma_\star}{H_\star} +4)},\eeq
where the $\star$ means that the quantity is evaluated at freeze out $c_s k/a(t_\star)\simeq \sqrt{\gamma_\star H_\star}$. In particular, in the strong dissipative regime,
 where $\gamma\gg H$, we obtain
\beq
\label{zeta2nu}
\Delta_{\zeta} \simeq {\dot f}^{\star2}\nu_{\calo}^\star \sqrt{\pi H_\star/\gamma_\star} \frac{H_\star^2}{2c_s^\star\left({c^\star_s}N_c^\star\right)^2}.
\eeq

Therefore, the tilt $n_s-1$ is given by
\begin{eqnarray} n_s-1&=&\frac{d \log \Delta_{\zeta} }{d \log k}\simeq\frac{1}{H^{\star}}\frac{d }{dt^{\star}}\log\left(\frac{\nu_{\calo}^\star{\dot f}^2_\star H^2_\star }{{N_c^\star}^2{c_s^\star}^3}\frac{16^{\frac{\gamma_\star}{H_\star} } (\frac{\gamma_\star}{H_\star} +1)^3
 \Gamma \left(\frac{\gamma_\star +H_\star}{2H_\star}\right)^4}{\pi  \Gamma ( \frac{2\gamma_\star}{H_\star} +4)}\right)\nonumber\\
&=&\epsilon_{{\nu}_{\cal{O}}}^\star+2\epsilon_f^\star-3 s^\star-2(\epsilon_{N_c}^\star+\epsilon^\star)+R\left(\gamma_\star/H_\star\right)(\epsilon^\star+\epsilon_{\gamma}^\star), \label{tilt}
\end{eqnarray} where we introduced the generalized slow roll parameters:
\beq \epsilon=-\frac{\dot{H}}{H^2},\,\, \epsilon_{\nu_{\calo}}=\frac{\dot{\nu}_{\cal{O}}}{H\dot{\nu}_{\cal{O}}},\,\, \epsilon_{f}=\frac{\ddot{f}}{H\dot{f}},\,\, \epsilon_{\gamma}=\frac{\dot{\gamma}}{2H\gamma},\,\, s=\frac{\dot{c_s}}{H c_s},\,\, \epsilon_{N_{c}}=\frac{\dot{N_c}}{H N_c}, \label{slowrollpar}
\eeq
and defined the function
\beq
R\left(x\right)=\frac{1}{2} x \left(-H_{\frac{x}{2}}+\frac{4}{x+1}+3 \psi\left(\frac{x+1}{2}\right)-2 \psi\left(x+\frac{5}{2}\right)+\gamma_E +\log
   (4)\right),\label{rfunct}
\eeq
with $H_{{x}}$ the Harmonic number, $\psi(x)$ the Digamma function and $\gamma_E$ the Euler's constant ($\gamma_E\simeq 0.577\ldots$). For $\gamma/H\gg 1$, the function $R$ behaves as
\be R\left(\frac{\gamma}{H}\right)\simeq -\frac{1}{2}+\frac{3}{8}\frac{H}{\gamma}+\calo\left(\frac{H^2}{\gamma^2}\right).\eeq 

Note we could have derived Eqs. (\ref{pow1gammazeta}) and (\ref{tilt}) working directly in terms of~$\zeta=-H\pi$. The~only difference would be the replacement $N_c \to {\tilde N}_c \equiv N_c/H$, and $\epsilon_{N_c} + \epsilon \to \epsilon_{{\tilde N}_c}$. One can also show the terms proportional to $\pi$ not displayed in Eq. (\ref{piaLO}), suppressed by generalized slow roll parameters, group themselves in such a way to transform the equation into an expression as in (\ref{piaLO}), but in terms of $\zeta = - H\pi$ with $N_c \to \tilde N_c$. We already studied some of these terms in the previous section, for more details see appendix \ref{constantz}.\\

In the ensuing sections we will demonstrate the validity of Eq. (\ref{consist}) at leading non-trivial order in the slow roll parameters and mixing with gravity scales. 

\section{The squeezed limit at first order in the generalized slow roll parameters} \label{sqz}

\subsection{Decoupling limit}

In order to set up the stage for the full computation including the mixing with gravity, here we start with the ingredients that enter in the consistency condition before adding corrections at $O(\epsilon)$, which will be the subject of the next section. In this limit we can work directly in the $\pi$-gauge, since $\zeta \simeq -H\pi$ with a constant $H$. 

 Even within this approximation the relationship in Eq. (\ref{consist}) becomes non-trivial. For example, notice in the equation for $\pi$ there is a term $\ddot f \pi\delta\calo$ (from expanding $f(t+\pi)$ to second order), whose contribution to $f_{\rm NL}$ in the squeezed limit is of order $\ddot f/(\dot f H) \sim \epsilon_f$ \cite{diseft}. 
 However, one can imagine a (somewhat tuned) situation in which $\epsilon_{\nu_\calo} + 2\epsilon_f \ll 1$ in Eq. (\ref{tilt}), namely $\dot f^2 \nu_\calo$ approximately time independent (to preserve scale invariance of the power spectrum) but with a larger $\epsilon_f$, in which case the consistency condition might not apply. As we show next that is not the case, and the resolution requires including contributions from the time dependence of the two-point function of the noise, i.e. $\nu_\calo(t)$.\\
 
We are interested in the three-point function of $\zeta$ in the squeezed limit. Hence, in addition to the non-linear term described above, and the non-linear interaction between $\pi$ and $\delta \calo^{(1)}$, we also need the response part $\delta \calo^s_R$ to second order in $\pi$. Since we are interested in computing the non-linear contributions 
to the equation for $\pi$ in the decoupling limit, namely ignoring terms proportional to $\dot{\pi}^2$ (and $(\partial\pi)^2$) which do not play a role in the squeezed limit when $\epsilon \to 0$, the only piece that contributes from Eq. (\ref{responseful}) comes from expanding $V_\calo(t+\pi)$, and takes the form 
\beq
\left(\delta\calo^s_R\right)^{(2)}(\pi) \simeq \dot{V}_\calo(t)\dot{\pi}\pi + \cdots.
\eeq  
Collecting all the ingredients the equation for $\pi$ to second order thus reads
\begin{eqnarray}
\ddot{\pi}_2+(3H+\gamma)\dot{\pi}_2-c_s^2\frac{\triangle\pi_2 }{a^2}\,&=&\,-\frac{\dot f}{N_c}\delta \calo_{\cal S}+\left(\epsilon_{N_c}-\epsilon_{f}\right)\frac{\dot{f}}{N_c}H\pi_1\delta\calo_{\cal S}+2 s{c_s^2}H\pi_1\frac{\triangle \pi_1}{a^2}\nn\\ && -\epsilon_{\gamma}H{\gamma}\pi_1\dot{\pi}_1,\label{piaNLO}\end{eqnarray} 
 where we only kept terms that contribute in the squeezed limit, and used Eq.~(\ref{piaLO}).\\

The  source term in Eq.~(\ref{piaNLO}) involving $\pi\delta \calo_{\cal S}$ was analyzed in \cite{diseft}. The details for the computations of $f_{\rm NL}^{\rm sq}$ for the remaining terms are given in appendix \ref{shapes}, and we find  
 \beq f_{\rm NL}^{\rm sq}\simeq-\frac{5 }{12}\left(2\epsilon_{f}+R{\left(\frac{\gamma}{H}\right)}{\epsilon_{\gamma}}-3 s-2\epsilon_{N_c}\right) \label{fnlddf},\eeq
with $R$ defined in Eq. (\ref{rfunct}). Therefore, even ignoring the mixing with gravity, up to this point the above expression shows that $f_{\rm NL}^{\rm sq}$ is proportional to the tilt if and only if $\nu_{\calo}$ is constant, namely $\epsilon_{\nu_{\calo}}=0$ in Eq. (\ref{tilt}). However, as we anticipated, the key observation is that this computation is incomplete, because at non-linear level there is also a contribution that arise from the fact that fluctuations in the equal time surfaces, described by $\pi$, affect the probability density functional for the noise when $\nu_\calo$ depends on time. 
 
To see how this contribution arises let us exploit the fact that the two-point function of the noise is a bi-scalar, hence we need to replace $\nu_{\calo}(t)\to\nu_{\calo}(t+\pi_1)$ in the unperturbed expression (\ref{twop1})  
\footnote{See appendix \ref{ctp} for a somewhat more formal discussion on the noise kernel.}
\beq
\label{noisemix0}
 \langle\delta\calo_{\cal S}(x_1)\delta\calo_{\cal S}(x_2)\rangle=\nu_{\calo}(t+\pi_1)\frac{\delta^4(x_1-x_2)}{\sqrt{-g}}\simeq \nu_{\calo}(t)\left( 1+\epsilon_{\nu_{\calo}} H\pi_1\right)\frac{\delta^4(x_1-x_2)}{a^3}.
\eeq 
This equation suggests (as shown in appendix \ref{ctp}) one can simply replace:\beq \delta\calo_{\cal S}\to\delta\calo_{\cal S}^{(1)} \left( 1+\frac{1}{2}\epsilon_{\nu_{\calo}} H\pi_1\right)\eeq 
in the equation for $\pi$, while keeping the unperturbed probability density functional for the noise. 
Therefore, the extra piece becomes equivalent to the term which is proportional to $\epsilon_f$, up to a factor of $1/2$. The upshot of this effect then boils down
 to replacing $\epsilon_f \to \frac{1}{2}\epsilon_{{\nu}_{\calo}}+\epsilon_{f}$ in Eq. (\ref{fnlddf}). This concludes the proof of the consistency condition in the decoupling limit, we study the mixing with gravity next.

\subsection{Including the mixing with gravity}\label{mix}

For computing the remaining contributions to $f_{\rm NL}^{\rm sq}$ that are suppressed by $O(\epsilon)$ we need to include the mixing with gravity. Here we will remain in the $\pi$-gauge. While working with $\zeta$ makes the physics more transparent, the main reason to use $\pi$ is that it allows us to proceed in complete generality without any particular knowledge about the action for the ADOF. That is not the case in the $\zeta$-gauge, since $S_{\calo}$ depends explicitly on the metric. The price to pay is perhaps a somewhat less evident conversion to physical quantities. This, however, does not represent a major issue while working perturbatively in the slow roll approximation.\\

First of all, notice that terms that involve short-wavelength modes $\delta{N}_S $ and $N^i_S$ do not contribute in the squeezed limit. For starters it is easy to show that terms containing $\delta N\pi$, namely without a derivative acting on $\pi$, are multiplied by a background function which is already of order one in the generalized slow-roll parameters. This is because factors of $\pi$ appear after expanding slowly varying functions of time. Since $\delta N$ is also order one in the slow roll parameters, or mixing scales, we can neglect all such terms (see Eq. (\ref{deltaN})). Furthermore we can also ignore quadratic terms containing  $\delta N_S \dot\pi_L$. Even though they are not multiplied by a generalized slow-roll parameter, in this case $\dot\pi_L$ yields an additional suppression (recall $\dot{\pi}_L \simeq \epsilon H\pi_L$).
However, terms involving $\delta N_L\dot \pi_S$ do produce a non-vanishing contribution to $f_{\rm NL}^{\rm sq}$ at leading order. Hence it is only necessary to account for $\delta N$ for long-wavelength modes. As we showed in sec. \ref{constr}, for the latter we have $\delta{N}_L\simeq \epsilon H\pi_L$.

Taking all these elements into account, expanding the action to cubic order and following the standard procedure, the equation for $\pi$ to second order becomes:
\begin{eqnarray}\label{piaNLO2}
&&\ddot{\pi}_2+(\gamma+3H)\dot{\pi}_2-c_s^2\frac{\triangle \pi_2}{a^2}=-\left(\epsilon_{N_c}-2\epsilon \right)H\pi_1(\ddot{\pi}_1+3H\dot\pi_1)\\
&&+(2 s+\epsilon_{N_c})c_s^2H\pi_1\frac{\triangle\pi_1}{a^2}-\frac{\dot{f}}{N_c}\left(\delta\calo^{(2)}_R+\delta\calo_{\cal S}\right)-\frac{\dot{f}}{N_c}\epsilon_fH\pi_1\left(\delta\calo^{(1)}_R+\delta\calo_{\cal S}\right)\nonumber\\
&&+\frac{4M_2^4}{N_c}(\dot{\pi}-\epsilon H\pi_1)\frac{\triangle\pi_1}{a^2}+\frac{6M_2^4}{N_c}3H(2\epsilon H\pi_1\dot{\pi}_1-\dot{\pi}_1^2)+\frac{12M_2^4}{N_c}\ddot{\pi}_1(\epsilon H\pi_1-\dot{\pi}_1)\nn \\ && +~{\rm terms~ that~do~not~contribute~in~the~squeezed~limit}.\nn
\end{eqnarray}

Recall that the relation between our $\pi$ and $\zeta$ is the same as in single field inflation (which is given in Eq. (\ref{relationzetapi})). However, as in the usual case, while in the long-wavelength limit a $\zeta$ mode remains constant, a $\pi$ mode does not. Hence a source term $\dot{\pi}^2_1$ will contribute to $f_{\rm NL}^{\rm sq}$. It is easy to see using  Eq. (\ref{relationzetapi}), that 
\be \dot{\pi}^2_1\simeq \frac{\dot{\zeta}^2_1}{H^2}+2\epsilon \zeta_1\frac{\dot{\zeta}_1}{H}+O(\epsilon^2)\simeq 2\epsilon H \pi_1\dot{\pi}_1+
O(\epsilon^2),\ee
and therefore, since $\dot\pi_L \simeq \epsilon H \pi_L$, the source terms in the last line of Eq. (\ref{piaNLO2}) do not contribute to $f_{\rm NL}^{\rm sq}$ at leading order.\\

In Eq. (\ref{piaNLO2}) we have both the response and noise parts. For the noise the two-point function is given by 
\beq
\label{noisemix}
 \langle\delta\calo_{\cal S}(x_1)\delta\calo_{\cal S}(x_2)\rangle=\nu_{\calo}(t+\pi_1)\frac{\delta^4(x_1-x_2)}{a^3 (1+\delta N)}\simeq \nu_{\calo}(t)\left(1+(\epsilon_{\nu_{\calo}}-\epsilon) H\pi_1\right)\frac{\delta^4(x_1-x_2)}{a^3},
\eeq
where we used Eq. (\ref{dneps}), and the fact that the two-point function for the noise is a bi-scalar. Whereas for the response we ought to use Eq. (\ref{responseful}), including now the metric components.  Expanding to the desired order we get 
\begin{eqnarray}
\left(\delta\calo^s_R\right)^{(1)}&\simeq& \frac{N_c\gamma}{\dot{f}}(\dot{\pi}-\delta N)\simeq \frac{N_c\gamma}{\dot{f}}(\dot{\pi}-\epsilon H\pi)+\cdots\label{Resp1}\\
\left(\delta\calo^s_R\right)^{(2)}&\simeq&\partial_t\left(\frac{N_c\gamma}{\dot{f}}\right)\pi\dot{\pi}+\frac{N_c\gamma}{2\dot{f}}\left(\dot{\pi}^2-4\dot{\pi}\delta N-\frac{(\partial_i\pi)^2}{a^2}\right)+W_\calo(t)\left(4\dot{\pi}^2-8\dot{\pi}\delta N\right)+\cdots \nonumber\\
&\simeq&\partial_t\left(\frac{N_c\gamma}{\dot{f}}\right)\pi\dot{\pi}-\frac{N_c\gamma}{\dot{f}}\epsilon H\pi\dot{\pi}+\cdots,\label{deltcal2}\end{eqnarray}
where  in the last line we used $\dot{\pi}^2\to 2 \epsilon H\dot{\pi}\pi+\cdots$, and we only kept terms that contribute in the squeezed limit. It is worthwhile emphasizing that, even though naively it seems to enter in the computation, the term proportional to $W_\calo(t)$ does not enter in the squeezed limit after we replace $\pi \to \pi_S+\pi_L$ and $\delta N_L \simeq \epsilon H \pi_L$, to leading order. This represents yet another check of the consistency condition, because had $W_\calo(t)$ contributed we would wind up with a term in the three-point function which is absent in the tilt of the power spectrum.\\

Adding all the pieces together the equation for $\pi$ then becomes
\begin{eqnarray}\label{piaNLO3}
&&\ddot{\pi}_2+\gamma\dot{\pi}_2-c_s^2\frac{\triangle \pi_2}{a^2}=-\frac{\dot f}{N_c}\delta \calo_{\cal S}+\left(\epsilon_{N_c}-\epsilon_f\right)\frac{\dot{f}}{N_c}H\pi_1\delta\calo_{\cal S}+2 s c_s^2 H\pi_1\frac{\triangle \pi_1}{a^2}-\epsilon_{\gamma}\gamma H\pi_1\dot{\pi}_1\nonumber\\
&&+\epsilon H\pi_1\left(-\gamma\dot{\pi}_1+2c_s^2\frac{\triangle\pi_1}{a^2}-2\frac{\dot{f}}{N_c}\delta\calo_{\cal S}\right),
\end{eqnarray} where we used the linearized equation for $\pi_1$, and wrote explicitly the response part $\delta\calo^s_R$ up to second order as shown above.

Using Eq. (\ref{piaNLO3}) we can compute the different contributions to the squeezed limit of the three-point function following the standard procedure. This is done in appendix \ref{shapes}. Finally we have to take into account the non-linear relation between $\zeta$ and $\pi$. Up to $O(\epsilon)$ it takes the form $\zeta\simeq-H\pi+\epsilon H^2\pi^2/2$, which corrects $f_{\rm NL}^{\rm sq}$ by a factor
\beq
 f_{\rm NL}^{\rm sq}{}_{\big|_{\rm nl}}\simeq \frac{5}{6}\epsilon. 
\eeq
   
Then, garnering all the contributions we obtain (see appendix \ref{shapes})
\beq
f_{\rm NL}^{\rm sq}\simeq-\frac{5}{12}\left(\epsilon_{{\nu}_{\cal{O}}}^\star+2\epsilon_f^\star-3 s^\star-2\epsilon_{N_c}^\star+R\left(\frac{\gamma_\star}{H_\star}\right)(\epsilon^\star+\epsilon_{\gamma}^\star)-2\epsilon^\star\right)= -\frac{5}{12}(n_s-1)\label{fnlsecmix}.
\eeq

One may be worried about the corresponding $\delta A^s_{\cal S}, \delta B^s_{\cal S}$ from the lapse and shift in Eq. (\ref{deltaN},\ref{deltaNi}), could in principle modify the computation of the three-point function in the squeezed limit. That is, however, not the case provided their two-point functions satisfy our approximations. The reason is twofold. First of all, they already enter suppressed by a power of $1/M_p^2$ in Eqs. (\ref{deltaN},\ref{deltaNi}). Moreover, since they also appear in the same way $\delta N$ does, terms like $\delta B^s_{\cal S} \pi$ will be equivalent to those of the form $\delta N \pi$, which as we argued before are already multiplied by a background function which is order one, making it higher order. Furthermore, since $\dot\pi_L \simeq \epsilon H \pi_L$, the only way they could contribute in the squeezed limit is via a coupling to $\dot \pi_S$. On the other hand, terms like $\delta B^s_{\cal S} \dot\pi_S$ can only contribute in the squeezed limit from the $\pi_L$ dependence in the noise two-point function, for example stemming from $\nu_B(t+\pi_L)$. However this will not contribute to the three-point function. (Also these are suppressed by slow roll parameters, as in Eq. (\ref{noisemix}), thus making their contribution a second order effect.)\\

This completes the proof of the consistency relation to leading non-trivial order in the generalized slow roll parameters. Furthermore, one can also show that the first correction to Eq. (\ref{fnlsecmix}) scales as $(k_L/k_S)^2$, for $k_L \to 0$. This is discussed in detail in appendix \ref{shapes} for the terms we studied in this paper, which contribute to the exact squeezed limit. For the remaining terms, such as those involving spatial derivatives of $\pi$, one can show using arguments similar to those displayed in \cite{cremi} that this scaling is not modified. (See appendix \ref{shapes}.) 

  \section{Rescaling away the long-wavelength mode}\label{genr}

In the standard interpretation of the consistency relation \cite{consist1,consist2,cremi,consist3}, a crucial point is the effect of long-wavelength modes $\zeta_L$ on the 2-point function for short scales at zeroth order in the gradients of $\zeta_L$, which reduces to a local rescaling of the coordinates $x^i_{\rm phys} = e^{\zeta_L} x^i$. In this section we show the same occurs in the presence of ADOF, however, for convenience we use $\pi$ rather than $\zeta$.\footnote{Even though, strictly speaking, $\pi_L$ is not conserved outside the horizon, our procedure is allowed to first order in the slow roll parameters, since $\dot\pi_L \simeq \epsilon H \pi_L$.} A long-wavelength mode $\pi_L$  considered as part of the background corresponds to a  deformation of the constant time surfaces of the form  $t \to \hat t-\pi_L$, so that now the perturbations are given by $\hat{\pi}=\pi-\pi_L\equiv \pi_S$.\footnote {Of course, by construction, the action is invariant under this transformation (provided we also transform the metric accordingly).} Then the metric in these coordinates takes the form 
\begin{eqnarray}
ds^2&=&-dt^2 (1+\delta N)^2+a^2(t)\delta_{ij}(dx^i+N^i dt)(dx^j+N^j dt)\nn\\
&=&-d\hat{t}^2 (1+\delta\hat{N})^2+a^2(\hat{t})\delta_{ij} e^{-2H\pi_L} (dx^i+\hat{N^i} d\hat{t})(dx^j+\hat{N^j} d\hat{t})\nn \\
&=&-d\hat{t}^2 (1+\delta\hat{N})^2+a^2(\hat{t})\delta_{ij}  (d\hat{x}^i+\hat{N}^i d\hat{t})(d\hat{x}^j+\hat{N}^j d\hat{t}),
\end{eqnarray} where we defined $\hat{x}^i = e^{-H\pi_L} x^i$. Also, to linear order in the perturbations,  $\delta\hat{N}=\delta{N}-\dot{\pi}_L$ and $\hat{N^i}={N^i}(1-\dot{\pi}_L)$. Then, since $\dot\pi_L \simeq \epsilon H\pi_L$ and $\delta N_L \simeq \epsilon H\pi_L$, we have \beq \delta \hat N_L \simeq 0,\eeq 
hence, to this order, the background metric written in terms of $(\hat{t},\hat{x}^i)$ takes the same form as in the original $\pi$-gauge. 
Therefore,  the equation for $\pi_S$ in the presence of $\pi_L$ is nothing but a re-scaling of the coordinates: $\hat t=t+\pi_L$ and  $\hat{x}^i = e^{-H\pi_L} x^i$ (or $\hat{k}^i = e^{H\pi_L} k^i$).\footnote {Note that the transformation $t\to t-\pi_L$, ${x}^i \to e^{-H\pi_L} x^i$ corresponds to an isometry of the background metric at zeroth order in the slow roll parameters.}
 
\noindent Expanding  Eq. (\ref{piaLO}) to linear order in $\pi_L$ we obtain:
\begin{eqnarray}\label{piashort}
\ddot{\pi}_S+(\gamma+3H)\dot{\pi}_S-c_s^2\frac{\triangle \pi_S}{a^2(t)}&=&-\frac{\dot{f}}{N_c}\delta\calo_{\cal S}+2\epsilon H\pi_L(\ddot{\pi}_S+3H\dot\pi_S)+2 sc_s^2H\pi_L\frac{\triangle\pi_S}{a^2}\\
&&+\frac{\dot{f}}{N_c}(\epsilon_{N_c}-\epsilon_f) H\pi_L\delta\calo_{\cal S}+(\epsilon-\epsilon_{\gamma})\gamma H \pi_L\dot{\pi}_S\nn \\
&=&-\frac{\dot f}{N_c}\delta \calo_{\cal S}+\left(\epsilon_{N_c}-\epsilon_f\right)\frac{\dot{f}}{N_c}H\pi_L\delta\calo_{\cal S}+2 s c_s^2 H\pi_L\frac{\triangle \pi_S}{a^2}\nn\\
&&-\epsilon_{\gamma}\gamma H\pi_L\dot{\pi}_S+\epsilon H\pi_L\left(-\gamma\dot{\pi}_S+2c_s^2\frac{\triangle\pi_S}{a^2}-2\frac{\dot{f}}{N_c}\delta\calo_{\cal S}\right),\nn
\end{eqnarray} where we used $\dot\pi_L \simeq \epsilon H \pi_L$. (To arrive at the second equality we used the linearized equation of $\pi_S$.) This is precisely the expression in Eq. (\ref{piaNLO3}).\\

To conclude our proof an important ingredient is the two-point function for the noise, which in the hatted coordinates is given by a similar expression to Eq. (\ref{twop1}), and therefore it transforms into:
\begin{eqnarray}
\langle \delta\calo_{\cal S}(\hat{t},{\hat{ \bf k}})\delta\calo_{\cal S}(\hat{t'},{\hat{\bf q}})\rangle& \simeq &\frac{\nu_{\calo}(\hat{t}) \delta(\hat{t}-\hat{t'})}{a^{3}(\hat{t})(1+\delta \hat N)} (2\pi)^3\delta^{(3)}({\hat{\bf q}+\hat{\bf k}})\nonumber\\
&\simeq& \nu_{\calo}(t)(1+(\epsilon_{\nu_{\calo}}-\epsilon) H\pi_L)\frac{ \delta({t}-{t'})}{a^3(t)} (2\pi)^3\delta^{(3)}({\bf q}+{\bf k})\label{noisepi},
\end{eqnarray}
up to leading order in $\pi_L$, and we used $\delta \hat N_L \simeq 0$. This again reproduces Eq. (\ref{noisemix}).

\section{Other type of couplings}\label{other}

In this paper we studied in detail the coupling $f(t)\calo$ for the interaction between $\pi$ and the ADOF. However, 
there are other possible terms one can write down. We review in what follows how to generalize our analysis to those cases.

\subsection{$\tilde\calo g^{00}$}

The coupling $-\frac{1}{2}\tilde\calo g^{00}$ was analyzed in detail in \cite{diseft}. At linear order, and assuming the time variation of the ADOF is much faster than the Hubble expansion, namely $H\delta{\tilde\calo} \ll \delta\dot{\tilde\calo}$,\footnote{These terms arise after integrating by parts the coupling $\int d^4x \sqrt{-g}a^3\tilde\calo\dot\pi$.} the only change with respect to the calculation in sec. \ref{stilt} is the replacement $\delta \calo \to \delta\dot{\tilde\calo}$, together with the requirement ${\rm Im} G^{\tilde\calo} \sim 1/\omega$, or $\delta\dot{\tilde\calo}^{(1)}_R \simeq \gamma N_c\dot{\pi}$ as in Eq. (\ref{ddff}), which guarantees the $\gamma\dot\pi$ dissipative term in the equation for $\pi$. 
Then the computation of the power spectrum and tilt follow the same steps. Note however that, by construction, in this case the time dependent function $f$ is not present, 
and therefore one should set $\epsilon_f \to 0$ and $\dot{f}\to 1$ in the previous calculations.

Under the local approximation, it is relatively straightforward to show that the results in secs. \ref{constr} and \ref{mix} still apply, modulo the above replacements. The manipulations, however, are somewhat different since the operator $-\frac{1}{2}\tilde\calo g^{00}$ contributes differently to the lapse and shift. Nevertheless, one can see that at leading order in the slow roll parameters the background part $\langle\tilde\calo\rangle$ can be absorbed into a {\it renormalization} of the $\bar\rho+\bar p$ term in Eq. (\ref{act1}). Therefore, under the same assumptions as in sec. \ref{constr}, one can show $\delta N \simeq \epsilon H \pi$ and the constancy of $\zeta$ holds.\\

For the direct computation of the three-point function the non-linear terms in the equation for $\pi$ are also not the same as before with a $f(t)\calo$ coupling. Moreover, because of the requirement ${\rm Im} G_{\tilde\calo} \sim 1/\omega$ \cite{diseft}, it is impossible to avoid a non-local behavior in some of the non-linear terms that appear in the
 equation for $\pi$, since under such assumption $\delta\tilde\calo$ does not have a fully local response to $\pi$. Similar terms also appear at linear (and non-linear) order in the response had we kept the sub-leading pieces proportional to $H\delta{\tilde\calo}$ in the equations. A proof that includes all such effects is beyond the 
scope of the present work, however, one can show that in the approximation where all such terms are neglected, that is under the assumption of a dominant local dynamics, the equation for $\pi$ takes a similar form to Eq. (\ref{piaNLO3}) (with $\epsilon_f=0$). The main observation is the following. After absorbing $\langle\tilde\calo\rangle$ into a redefinition of  $\bar\rho+\bar p$, at non-linear order the contribution from $\delta\tilde O$ to the equation for $\pi$ can be written as
\begin{eqnarray}
\frac{1}{\sqrt{-g}}\frac{\delta \left(-\frac{1}{2}\int d^4x \sqrt{-g}\delta \tilde\calo g^{00}(\pi)\right)}{\delta \pi}&=&\frac{\partial_{\mu}\left(\sqrt{-g}\delta\tilde\calo g^{\mu\nu}\partial_{\nu}(t+\pi)\right)}{\sqrt{-g}} \nn \\
&\simeq& g^{\mu\nu}\partial_{\nu}(t+\pi)\partial_{\mu}\delta\tilde\calo \nonumber\\
&=& -\sqrt{-g^{\rho\nu}\partial_{\rho}(t+\pi)\partial_{\nu}(t+\pi)}\,\, n^{\mu}\partial_{\mu}\delta\tilde{\calo},
\end{eqnarray}
where $n^\mu$ is defined in Eq. (\ref{nmu}), and in the second line we kept only derivatives of $\delta\tilde\calo$, as explained above. Using this expression, and expanding up to second order we obtain on the RHS of the equation for $\pi$ the terms:
\beq
-\frac{1}{N_c}(1+\dot{\pi}-\delta N) \left(n^{\mu}\partial_{\mu}\tilde{\calo}\right)^{(1)}-\frac{1}{N_c} \left(n^{\mu}\partial_{\mu}\tilde{\calo}\right)^{(2)}
\eeq
where we wrote only the terms that may contribute in the squeezed limit and made explicit the appearance of the scalar operator: $\calo \equiv n^\mu \partial_\mu\tilde\calo$. Then, we notice that the only difference with respect to our analysis in sec. \ref{mix} is the term: $(\dot{\pi}-\delta N) \left(n^{\mu}\partial_{\mu}\tilde{\calo}\right)^{(1)}$. On the other hand, recall $\dot\pi_L-\delta N_L \simeq 0$, therefore this piece does not contribute to the squeezed limit at the order we work here when we consider the stochastic part, namely $\left(n^{\mu}\partial_{\mu}\delta\tilde{\calo}\right)^{(1)}_{\cal S}$, given by $\delta\dot{\tilde\calo}_{\cal S}$. However, it may still contribute when instead we take the long-wavelength limit for the response part of the operator and evaluate $\pi$ on the short mode. But, as in Eqs. (\ref{resps},\ref{Resp1}), we have \beq \left(n^{\mu}\partial_{\mu}\tilde{\calo}\right)^{(1)}\simeq \gamma N_c(\dot{\pi}-\delta N),\eeq which as we argued does not enter in $f_{\rm NL}^{\rm sq}$ when evaluated for long-wavelength modes. (Note, nonetheless, this term produces the $\gamma\dot\pi$ dissipative correction for short modes.) This concludes the proof of the consistency condition for this type of coupling, under the assumption of a dominant local interaction.

\subsection{Vectors \& Tensors}

For vector and tensor operators the story changes slightly. For example, we can have a coupling $\calo_\mu g^{\mu0}$, which may also produce a $\gamma\dot\pi$ dissipative term. The difference with the $\tilde \calo g^{00}$ term is that the former introduces only a linear coupling between $\partial_\mu \pi$ and $\calo_\mu$ at the level of the action. Again, under the same assumptions as in the previous subsection, the computation of the power spectrum and proof of the consistency condition follow similar steps as in secs. \ref{stilt} and \ref{mix}, after the identification: 
\beq 
\calo \to \frac{1}{\sqrt{-g}}\partial_{\mu}(\sqrt{-g}{\calo}^{\mu}).
\eeq
Similar considerations apply for tensor couplings.

\section{Conclusions}

Up until now, checks of the consistency condition had been limited to single field inflation. As it is well known, multi-field models with many light degrees of freedom relevant 
during inflation are capable of violating such condition \cite{multieft}. In this paper we filled a gap in the literature and showed the validity of the consistency relation for a vast class of (multi-field) models having a preferred clock. More specifically, using the EFT framework developed in \cite{diseft}, we explicitly demonstrated that:
\begin{eqnarray}
\label{last}
\lim_{k_1\to 0} \langle\zeta_{k_1}\zeta_{k_2}\zeta_{k_3}\rangle=-(2\pi)^3\delta^3\left(\sum_i {\bf k}_i\right)P_{\zeta}(k_L)P_{\zeta}(k_S)(n_s-1),
\end{eqnarray}
holds for dissipative single-clock inflation, to first non-trivial order in the generalized slow roll parameters and mixing with gravity scales. Moreover, we also showed that the first correction to the exact squeezed limit scales as $(k_L/k_S)^2$ when $k_L \to 0$.
Unlike cases where the curvature perturbations are produced in the Bunch-Davies vacuum, here the proper treatment of the contribution from the noise to the power spectrum and non-Gaussianities played a key role.\\

Following \cite{diseft} our main assumptions were, in addition to the existence of a preferred clock, the validity of the local approximation, derivative expansion, and the emergence of a shift symmetry at the level of the response for the ADOF. From here we were able to show, after including the mixing with gravity, that the effects from the ADOF turn off while $\zeta$ remains constant outside the horizon. The latter was crucial, in particular for long-wavelength modes, in allowing us to prove $\delta N_L \simeq \epsilon H\pi_L$ in the $\pi$-gauge, which resembles what occurs in single field models. Given the generality of our EFT approach \cite{diseft} we relied on the (non-linear) realization of the symmetries, which we exploited extensively in order to incorporate the couplings to $\pi$, as well as the mixing with gravity. Another advantage of the EFT formalism, and working with $\pi$ even in the long-wavelength limit (where the meaning of $\pi$ is somewhat less transparent), is that it allows us to bypass any specific knowledge about the dynamics of $\calo$, other than the analytic properties of Green's function and stochastic noise. This would not be the case in the unitary gauge, since $\zeta$ couples to the ADOF in a completely unknown manner in $S_{\calo}$.\\

The results in this paper apply to a plethora of possible scenarios with ADOF, including warm \cite{warm} and trapped inflation\footnote{Technically speaking trapped inflation corresponds to a series of operators of the form $\sum_i f_i\calo_i$. It is easy to show the results of this paper apply to this case as well.} \cite{trapped}, and provide further support to the claim that the squeezed limit of the three-point function offers a remarkable opportunity to probe the very mechanism behind primordial density fluctuations.\footnote{Other soft limits on $n$-point functions have been recently studied in \cite{paolo,lam,dd4}.} The forthcoming results from the PLANCK satellite \cite{planck}, as well as large scale structure measurements \cite{sqz1,sqz2} and CMB $\mu$-distortion \cite{mud}, thus have the ability to significantly constrain, and/or rule out, a large(r) class of inflationary models.

\begin{center}
{\bf Acknowledgements}
\end{center}
We thank Leonardo Senatore for helpful discussions. This work was supported by: Universidad de Buenos Aires, CONICET and ANPCyT (DLN); NSF grant AST-0807444 and DOE grant DE-FG02-90ER40542 (RAP); NSF grants PHY-0855425, AST-0506556 \& AST-0907969, and by the David \& Lucile Packard and the John D. \& Catherine T. MacArthur Foundations (MZ). 
\appendix

\section{$\zeta_L$ in the slow roll approximation}\label{constantz}

We want to show that the equation for $\zeta$ resembles Eq. (\ref{piaLO}) to first order in the slow roll approximation, including the mixing with gravity. Let us ignore the noise part for simplicity. Let us also re-write the background equations in (\ref{beqmet})-(\ref{bacO}) one more time:
\begin{eqnarray}
&& 3H^2 M_p^2=\bar{\rho}+\bar{\rho}_{\calo}+f(t)\bar\calo, \label{beqmet2}\\
&&\dot{\bar\rho}+3H(\bar\rho+\bar p)+\dot{f}\bar{\calo}=0,\label{beqpi2}\\
&&\dot{\bar\rho}_{\calo}+3H(\bar\rho_{\calo}+\bar p_{\calo})+f\dot{\bar{\calo}}=0.\label{bacO2}
\end{eqnarray} 

Expanding the full action the to quadratic order, the linearized equation for $\pi$ (to all order in slow-roll)~is~given~by
\begin{eqnarray}
\frac{1}{Na^3}\left(\frac{\delta S}{\delta \pi}\right)^{(1)}&=&-N_c(\ddot{\pi}-\dot{\delta N})-(\dot{N}_c+3HN_c)(\dot{\pi}-\delta{N})-[\ddot{\bar \rho}+3H(\dot{\bar p}+\dot{\bar \rho})]\pi+3H(\bar{p}+\bar{\rho})\delta N\nonumber\\
&-&\dot{f}(\calo-\bar{\calo})-\ddot{f}\pi \bar{\calo}+\ldots,
\end{eqnarray} where the dots stand for terms that contain spatial derivatives and $N^i$ which we ignore since we are interested in the long-wavelength limit where $k_L \to 0$. (Since, as discussed in sec. \ref{constr}, the ADOF do not affect the lapse and shift it is easy to show that the terms that survive in the long-wavelength limit are proportional to $\dot \zeta$, and moreover suppressed by factors of $\epsilon$ \cite{maldacena,consist3}.) \\

Let us start at first order in the slow roll parameters. Note that the equation for $\pi$ has a term which does not involve derivatives, this is required for $\zeta = - H \pi$ to be massless. Using the derivative of Eq. (\ref{beqpi2}), $\delta N_L\simeq \epsilon H\pi_L$ together with $\dot{\pi}_L\simeq \epsilon H\pi_L$, we find (ignoring gradients): 
\beq
\frac{1}{Na^3\tilde N_c}\left(\frac{\delta S}{\delta \pi}\right)^{(1)}= \ddot \zeta + 3H \dot \zeta -\frac{\dot{f}}{\tilde N_c}(\calo-\bar{\calo}-\dot{\bar{\calo}}\pi)=\ddot \zeta + 3H \dot \zeta -\frac{\dot{f}}{\tilde N_c}\left(\delta\calo^s_R\right)^{(1)}(\pi),
\eeq 
up to second order in slow roll, with $\tilde N_c = N_c/H$. Hence $\zeta$ will be conserved outside the horizon because 
\beq
\frac{\dot{f}}{\tilde N_c}\left(\delta\calo^s_R\right)^{(1)}(\pi_L) \simeq \gamma H \left(\dot \pi_L - \epsilon H \pi_L\right) = -\gamma \dot \zeta_L,
\eeq
for $k_L\to 0$, which is the case as we argued in sec. \ref{constr}. 

We can also demonstrate that the above reasoning applies also to second order in slow roll. The main difference, once again, is the appearance of  $\left(\delta\calo^s_R\right)^{(2)}=\calo-\bar{\calo}-\dot{\bar{\calo}}\pi-\ddot{\bar{\calo}}\pi^2/2$, which is then required to be proportional to derivatives of $\zeta$ in the long-wavelength limit.  Moreover, all other terms pair up such that the resulting equation for $\zeta_L$ does not have any contribution which does not involve derivatives.

\section{The source terms in the squeezed limit \& $(k_L/k_S)^2$ scaling}\label{shapes}

In this appendix we provide some details on the computation of $f_{\rm NL}^{\rm sq}$ for the different types of source terms appearing in  Eqs. (\ref{piaNLO}) and (\ref{piaNLO3}). We also show that the first non-trivial correction in the squeezed limit scales as $(k_L/k_S)^2$.\\

Let us start with the source term that is proportional to $\epsilon_f$: $-N_c^{-1}\dot{f}\epsilon_{f}H\pi\delta\calo_{\cal S}$. The contributions of the other terms involving $\pi\delta\calo_{\cal S}$ can be obtained from here by performing the appropriate replacements. To analyze the non-Gaussianities we  decompose $\pi=\pi_1+\pi_2$, where again the subscripts represent the order of the solution for the fluctuations. Then
\be\label{pi1nl}\pi_1(k,\eta)=\frac{\dot{f} k c_s}{N_cH^2}\int^{\eta}_{\eta_0} d\eta' g_{\gamma}(k c_s|\eta|,k c_s|\eta'|) \delta{ \calo}_S,\ee where
$g_{\gamma}(k c_s|\eta|,k c_s|\eta'|)=G_{\gamma}(k c_s|\eta|,k c_s|\eta'|)/(k c_s{\eta'})^2$ with $G_{\gamma}$ defined in Eq. (\ref{greengamma}),
and (using Eq. (\ref{pi1nl}))
\begin{eqnarray}
\pi_2(k_3,0)&=&\frac{\epsilon_f  \dot{f}^2k_3 c_s^2}{ N_c^2H^3} \int^0_{\eta_0} d\eta'   g_{\gamma}(0,k_3c_s|\eta'|)\int\frac{d^3{\bf q}}{(2\pi)^3} q  \nonumber\\
&\times&
\int^{\eta'}_{\eta_0} d\eta'' g_{\gamma}(qc_s|\eta'|,qc_s|\eta''|) \delta{\calo}_S({\bf q},\eta'')\delta{\calo}_S({\bf k}_3-{\bf q},\eta').\label{pi2nl}
\end{eqnarray}

We want to compute the three-point function for $\zeta\simeq -H\pi$ in the limit $\eta\to 0$:
\begin{eqnarray}\langle\zeta({\bf k}_1,0)\zeta({\bf k}_2,0)\zeta({\bf k}_3,0)\rangle&=&\langle\zeta_1({\bf k}_1,0)\zeta_1({\bf k}_2,0)\zeta_2({\bf k}_3,0)
\rangle+ \mbox{cyclic sum in $k_i$'s}\nonumber\\
&=&(2\pi)^3 \delta^3(\sum_i{\bf k}_i) F(k_1,k_2,k_3). \end{eqnarray}
Then, using  Eq. (\ref{twop1}) and that at leading order the noise is Gaussian we have
\begin{eqnarray}\label{os}&&\langle\delta{ \calo}_S({\bf k}_2,\tilde{\eta'}) \delta{\calo}_S({\bf k}_1,\tilde{\eta})  \delta{\calo}_S({\bf q},\eta'')
\delta{\calo}_S({\bf k}_3-{\bf q},\eta''')\rangle=(2\pi)^6\nu_{{\calo}}^2\frac{\delta^{(3)}({\bf k}_1+{\bf k}_2+{\bf k}_3)}{a^4(\eta'')a^4(\eta''')}\\
&&\times \left\{\delta(\tilde{\eta'}-\eta'')\delta(\tilde{\eta}-\eta''')\delta^{(3)}({\bf k}_2+{\bf q})+
\delta(\tilde{\eta}-\eta'')\delta(\tilde{\eta'}-\eta''')\delta^{(3)}({\bf k}_1+{\bf q})\right\}\,\,\,(\mbox{for} \,\,\,k_3\neq 0)\nn.
\end{eqnarray}
Defining $x_i=k_i/k$, with $k$ an arbitrary scale with units of momentum, we obtain (for $\eta_0\to -\infty$):
\begin{eqnarray}\label{fdeltashape}
&&F(x_1,x_2,x_3)= k^6 F(k_1,k_2,k_3)=-\frac{ H^4\nu_{{\calo}}^2\dot{f}^4\epsilon_{f}}{N_c^4c_s^6}x_1x_2^2x_3\int_{0}^{+\infty} dy\, y^4  g_{\gamma}(0,x_3 y)g_{\gamma}(0,x_1 y)\nonumber\\
&&\times\int_{y}^{+\infty} dz\,{z}^4 \,g_{\gamma}(x_2 y,x_2 z)  g_{\gamma}(0,x_2 z)+\mbox{permutations  in  $x_i$'s}
\end{eqnarray} (we performed a change of variables: $y'=-kc_s\eta'$, $z=-kc_s\eta''$).

From the definition of $f_{\rm NL}^{\rm sq}$ given in Eq. (\ref{squ}) and the expression in Eq. (\ref{pow1gammazeta}) for the power spectrum, it is then straightforward to show
\be f_{\rm NL}^{\rm sq}{}_{\big|_{\epsilon_f}}=-\frac{5}{6}\epsilon_f,\ee
independently of the value of $\gamma$.\\

We follow now a similar procedure for the contribution due to the change in the probability density functional of the noise. In this case we need
\begin{eqnarray}
&&\langle\zeta({\bf k}_1,0)\zeta({\bf k}_2,0)\zeta({\bf k}_3,0)\rangle=-\frac{\dot{f}^3c_s^3k_1k_2k_3}{N_c^3H^3}\int_{-\infty}^0 d\eta_1\int_{-\infty}^0 d\eta_2\int_{-\infty}^0 d\eta_3  g_{\gamma}(0,c_sk_1|\eta_1|)\\
&&\times g_{\gamma}(0,c_sk_2|\eta_2|) g_{\gamma}(0,c_sk_3|\eta_3|) \langle\delta\calo_{\cal S}(\eta_1,{\bf k}_1)\delta\calo_{\cal S}(\eta_2,{\bf k}_2)\delta\calo_{\cal S}(\eta_3,{\bf k}_3)\rangle.\nn
\end{eqnarray} Using Eq. (\ref{cont}) and
\begin{eqnarray}\label{os6}&&\langle\delta{ \calo}_S({\bf k}_1,\eta_1) \delta{\calo}_S({\bf k}_2,\eta_2)  \delta{\calo}_S({\bf k}_3,\eta_3)
\delta{\calo}_S({\bf k}, \eta')\delta{\calo}_S({\bf p}, \eta')\delta{\calo}_S({\bf -k-p}, \eta'')\rangle=\\ 
&=&(2\pi)^9\nu_{\calo}^3\frac{\delta^{(3)}({\bf k}_1+{\bf k}_2+{\bf k}_3)}{a^4(\eta_3)a^4(\eta')a^4(\eta'')}\left\{\delta({\eta}_1-\eta')\delta({\eta}_2-\eta'')\delta({\eta}_3-\eta')\delta^{(3)}({\bf k}_1+{\bf k})\delta^{(3)}({\bf k}_3+{\bf p})\right.\nn\\  &+& \left. \mbox{permutations in the $k_i$'s}\right\},\nn
\end{eqnarray} one can show that this term yields a contribution to $F(x_1,x_2,x_3)$ that is exactly the same as the one in Eq. (\ref{fdeltashape}) after
 replacing $\epsilon_{f}$ by $\epsilon_{{\nu}_{\calo}}/2$.\\

Let us now analyze the contribution  to $f_{\rm NL}^{\rm sq}$ from the last term on the RHS of Eq.(\ref{piaNLO}): $-\epsilon_{\gamma}H{\gamma}\dot{\pi}\pi$. 
This term contains a time derivative of $\pi$, but it also involves an additional temporal integration. 
For this reason it is non-trivial to see whether this vanishes or not in the squeezed limit. In this case we have
\begin{eqnarray}
F(x_1,x_2,x_3)&=&\frac{\epsilon_{\gamma} H^3\gamma\nu_{\calo}^2\dot{f}^4 }{N_c^4c_s^6} x_1 x_2^2 x_3^2\int_0^{+\infty} dy\int_y^{+\infty} dz \int_y^{+\infty} dw \,y z^4 w^4 g_{\gamma}(0,x_1y)\\
&&g_{\gamma}(0,x_3z) g_{\gamma}(x_3y,x_3z)g_{\gamma}(0,x_2w) \frac{d}{dy}g_{\gamma}(x_2y,x_2w)+\mbox{permutations  in  $x_i$'s}.\nn
\end{eqnarray}

In the squeezed limit $x_3\to 0$, $x_1\to x_2 =1$, the contribution that does not vanish is given by
\begin{eqnarray}
\label{cccc}
C\equiv\lim_{x_3\to 0} x_3^3 F(1,1,x_3)&=& \lim_{x_3\to 0} \frac{2\epsilon_{\gamma} H^3\gamma \nu_{\calo}^2\dot{f}^4}{ N_c^4c_s^6} \int_0^{+\infty} d\tilde{y}\int_{\tilde{y}x_3}^{+\infty} d\tilde{z}
 \int_{\tilde{y}}^{+\infty} d\tilde{w} \,\tilde{y} \tilde{z}^4 \tilde{w}^4 g_{\gamma}(0,\tilde{y}) \nonumber\\
&&g_{\gamma}(0,\tilde{z}) g_{\gamma}(x_3\tilde{y},\tilde{z})g_{\gamma}(0,\tilde{w}) \frac{d}{d\tilde{y}}g_{\gamma}(\tilde{y},\tilde{w})\nonumber\\
&=& \frac{2\epsilon_{\gamma} H^3\gamma \nu_{\calo}^2\dot{f}^4}{ N_c^4c_s^6} \int_0^{+\infty} d\tilde{y}\int_{0}^{+\infty} d\tilde{z}
 \int_{\tilde{y}}^{+\infty} d\tilde{w} \,\tilde{y} \tilde{z}^4 \tilde{w}^4 g_{\gamma}(0,\tilde{y}) \nonumber\\
&&g_{\gamma}(0,\tilde{z}) g_{\gamma}(0,\tilde{z})g_{\gamma}(0,\tilde{w}) \frac{d}{d\tilde{y}}g_{\gamma}(\tilde{y},\tilde{w})
\end{eqnarray} where we made the change of variables $\tilde{y}=x_1y$, $\tilde{z}=x_3z$, $\tilde{w}=x_2w$ and we set $x_1=x_2=1$. The contribution to $f^{\rm sq}_{\rm NL}$ reads 
\be \label{rtilde}
 f_{\rm NL}^{\rm sq}{}_{\big|_{\epsilon_\gamma}}=\frac{5}{12}\epsilon_{\gamma} \frac{C}{P_{\zeta}^2(1)} = -\frac{5}{12} \epsilon_{\gamma}R\left(\frac{\gamma}{H}\right),
\ee with $R(\gamma/H)$ given in Eq. (\ref{rfunct}).\footnote{Even though this equality is not straightforward, one can show that the integrand in Eq. (\ref{cccc}) agrees with an equivalent representation of Eq. (\ref{rfunct}) term by term in an analytic expansion, once the common non-analytic pieces are factored out. Moreover, the results also agree numerically.} 

Similarly we obtain a piece proportional to $\epsilon$ from Eq. (\ref{piaNLO3}) after replacing $\epsilon_\gamma \to \epsilon$, and for the source term $2 s{c_s}H\pi_1\frac{\triangle\pi_1}{a^2}$ we find: $f_{\rm NL}^{\rm sq}{}_{\big|_s}= \frac{5}{4}s$, independently of $\gamma$. \\

To probe the scaling with $k_L/k_S$ from each one of these terms it is sufficient to take each contribution and expand in powers of $x_3$. For suggestive
 purposes in what follows we re-label $x_3 \to x_L$. To show that the first non-trivial correction starts at $O(x_L^2)$ we just need to show that the first
 derivative of $x_L^3F(x_S,x_S,x_L)$ vanishes at $x_L=0$. Here we assume analyticity of the result in $x_L$. (This is not required by the consistency condition, since for instance we could still have $\sqrt{x_L}$. However, by inspection it is easy to show each contribution is indeed analytic.) \\

As an example, let us consider the term in Eq. (\ref{cccc}) but for non-zero $x_L$. If we take now ${\partial C(x_L) \over \partial x_L}$, we encounter a few different contributions. First of all we have the piece that depends on the limit of integration, and therefore we need to evaluate the integrand at $x_L$. This produces pieces proportional to $g_\gamma(0,0)$ which vanish.
The remaining term depends on the integral of a derivative of the Green's function, for example $\partial_{x_L} g_\gamma(x_L y,z)$. Then, using the property \cite{diseft}
\beq
\partial_{x_L}g_{\gamma}(x_L y, z)\sim \frac{x_L}{(1+\gamma/H)} y^2 g_{\gamma}(x_L y,z),
\eeq  
which is valid for $y x_L \ll y^\star \simeq \sqrt{\gamma/H}$, and taking into account that the contribution from $g_\gamma(0,y)$ in the integral is dominated for values $y \simeq y^\star$,\footnote{See appendix F in \cite{diseft}} one concludes that this term scales like $x_L$ for $x_L \ll 1$, hence ${\partial C \over \partial x_L}$ vanishes when $x_L\to 0$ as advertised.\\

One may wonder whether the terms we ignored in this paper, that do not contribute in the exact squeezed limit, could in principle produce a correction at order $k_L/k_S$. As shown in \cite{diseft} that is not the case for the pieces proportional to $\delta \calo \dot \pi$, since in fact they are sub-dominant in the squeezed limit. Moreover, following \cite{cremi}, one can also show that the other possible terms studied in \cite{diseft}, such as $\gamma (\partial\pi)^2$ etc., do not modify the scaling in Eq. (\ref{consist}). The basic idea is the following. Since the lapse and shift are determined by the same expressions as in single field models, the arguments put forward in \cite{cremi} translate directly to our case. For example, one can have a term of the form $N^i \partial_i \pi \delta\calo$, from the mixing with gravity. But $\delta N \simeq O(k_L^2)$ and $N^i \simeq O(k_L)$, and therefore there is no left over $k_L/k_S$ upon symmetrization \cite{cremi}. Note also that terms coming from the vector coupling $\calo^\mu \partial_\mu \pi$ turn into a scalar interaction of the form $(\partial_\mu \calo^\mu) \pi$. Moreover, couplings of the sort $g^{\mu0}\calo_\mu$ do not produce non-linear terms in $\pi$ other than through the non-linear response for $\partial_\mu \calo^\mu$, which is included in our analysis.

\section{Non-linear generalization of the  probability density functional for the noise }\label{ctp}

The issue of including the effect of  $\pi$ on the probability density functional for the noise
can be naturally addressed in the framework  of the closed time path (CTP) or Schwinger-Keldysh
 formalism, where 
one has a CTP effective action for the  ``mean''  field, that is given by $\pi_{+}=(\pi^1+\pi^2)/2$, and the difference $\pi_{-}=\pi^1-\pi^2$ (in CTP notation).
In general, the effective action has  real  and  imaginary parts (see for instance \cite{calzetta}),
\be \exp\{-i\Gamma[\pi_{+},\pi_{-}]\}=\exp\{-i \Re{\Gamma}[\pi_{+},\pi_{-}] +\Im{\Gamma}[\pi_{+},\pi_{-}]\}.\ee
The imaginary part  can be interpreted as the result of an averaging over an stochastic source $\xi\equiv\delta\calo_{\cal S}$.
 For example, for the coupling $f(t)\calo$ and to second order in $\pi$, the imaginary part of the effective action takes the form
\begin{eqnarray}
&&\exp\{\Im{\Gamma\}[\pi_{+},\pi_{-}]}=F[\pi_{-}(x)]\nonumber\\
&&=\exp\left\{-\frac{1}{2}\int^t d^4x_1\, a^3(t_1)\int^t d^4x_2 \,a^3(t_2)\;\Delta_{-}(x_1)N(x_1,x_2)\Delta_{-}(x_2)\right\},
\end{eqnarray}  with $\Delta_{-}=\dot{f}\pi^2+\ddot{f}(\pi^2)^2/2-\dot{f}\pi^1-\ddot{f}(\pi^1)^2/2$, and $N(x_1,x_2)$ is the noise kernel. The probability density functional of the stochastic source is then obtained after using the following mathematical identity
\be \label{eeeF}
F[\pi_{-}(x)]=\int\mathcal{D}\xi(x)P[\xi(x)]\exp\left\{-{i}\int^{t}d^4y\,a^3\;\Delta_{-}(y)\xi(y)\right\},
\ee  
with 
\begin{equation}\label{ProbRuido}
P[\xi]=\mathcal{A}\exp\left\{-\frac{1}{2}\int^{t}d^4{x_1}\,a^3(t_1)\int^{t}d^4{x_2}\,a^3(t_2)\:\xi(x_1)N^{-1}(x_1,x_2)\xi(x_2)\right\},
\end{equation} where $\mathcal{A}$ is a  normalization constant and $N^{-1}(x_1,x_2)$ is the inverse functional of $N(x_1,x_2)$.
In the local approximation (in our case)
\begin{eqnarray} N(x_1,x_2)&=&\nu_{\calo}(t)\frac{\delta^4(x_1-x_2)}{a^3} \label{noisekernel}\\
 N^{-1}(x_1,x_2)&=&\frac{1}{\nu_{\calo}(t)}\frac{\delta^4(x_1-x_2)}{a^3}.
\end{eqnarray}

The equation for $\pi$, as an stochastic variable, is then obtained before averaging over $\xi$, namely from the real part of the action where we have the term $\int^{t}d^4y\,a^3\;\Delta_{-}(y)\xi(y)$.\footnote{The variation can be taken either with respect to $\pi^1$ or $\pi^2$, and afterwards one sets $\pi^1=\pi^2$ (i.e., $\pi_{+}=\pi$ and $\pi_{-}=0$).} The contribution for this term is equivalent to our $\dot{f}\delta\calo_{\cal S}+\dot{f}\epsilon_f H\pi\delta\calo_{\cal S}$ in Eq. (\ref{piaNLO}).

In cases where the probability distribution depends on $\pi_{+}$ (i.e., $P[\xi]\to P[\xi,\pi_{+}]$) the previous identity remains valid. This justifies the replacement in the noise kernel $\nu_{\calo}\to \nu_{\calo}(t+\pi)$ (recall that at the level of the equation of motion $\pi_{+}=\pi$).\\

Our objective is to compute $\langle\pi\pi\pi\rangle$ perturbatively. So, we split $\pi=\pi_{1}+\pi_{2}$, where $\pi_{2}$ (like  $\pi_{1}$) has $\delta\calo_{\cal S}=\xi$ as a source, but for $\pi_2$  
the averaging over the noise $\langle\ldots \rangle$ is computed using $P[\xi,\pi_{1}]$, which is obtained after the replacement  $\nu_{\calo}\to\nu_{\calo}+\dot{\nu}_{\calo}\pi_{1}$. Therefore, we need to know 
 $\langle\xi\xi\xi\rangle$ obtained using $P[\xi,\pi_{1}]$:
\be\label{cont}
\langle \xi\xi\xi\rangle \simeq\int \mathcal{D}\xi P[\xi,\pi_{+}=0]\left(\frac{1}{2}\int\int d^4x_1 a^3(t_1) d^4x_2 a^3(t_2)\xi N^{-1}\xi)\frac{\dot{\nu}_{\calo}}{{\nu_{\calo}}} \pi_{1}\right) \xi\xi\xi\Big{|}_{\pi_{-}=0}.\ee

Now we have to replace $\pi_1$ by the solution in Eq. (\ref{pi1}). Note that there are six stochastic fields ($\xi$'s) in Eq. (\ref{cont}) but in the denominator there are two factors of $\nu_{\cal{O}}$ (one is inside $N$) and we have a factor $1/2$. It is not difficult to see that the result is exactly the
 same as the one for $-N_c^{-1}\dot{f}\epsilon_f H\pi_1\delta\calo_{\cal S}$ (see appendix \ref{shapes}), but replacing $\epsilon_f\to\epsilon_{{\nu}_{\cal{O}}}/2$. Then, the effect of both contributions becomes 
\beq  f_{\rm NL}^{\rm sq} {}_{\big|_{(\epsilon_f,\epsilon_{\nu_\calo})}}\simeq-\frac{5 }{12}\left(\epsilon_{{\nu}_{\cal{O}}}+2\epsilon_{f}\right).\eeq

Perhaps a more straightforward way to show that this is the combination that appears in the three-point function is to use $\Delta_{-}=\dot{f}\pi_{-}+\ddot{f}\pi_{-}\pi_{+}$, and rewrite Eq.(\ref{eeeF}) as
\be \label{eeeFBis}
F[\pi_{-}(x)]=\int\mathcal{D}\xi(x)\tilde{P}[\xi(x)]\exp\left\{-{i}\int^{t}d^4ya^3\;\dot{f}\pi_{-}(y)\xi(y)\right\},
\ee  
where in the equation for $\pi_2$ (Eq. (\ref{piaNLO})) we replace the source term $\dot{f}\delta\calo_{\cal S}+\dot{f}\epsilon_fH\pi_1\delta\calo_{\cal S}$ by $\dot{f}\tilde{\delta\calo_{\cal S}}$, but with a different probability density functional:
\begin{equation}\label{ProbRuidobis}
\tilde{P}[\xi]=A\exp\left\{-\frac{1}{2}\int^{t}d^4{x_1}\,a^3(t_1)\int^{t}d^4{x_2}a^3(t_2)\:\xi(x_1)\hat{N}^{-1}(x_1,x_2)\xi(x_2)\right\},
\end{equation} with
\begin{eqnarray}  
\hat{N}^{-1}(x_1,x_2)&=&(\nu_{\calo}+\dot{\nu}_{\calo}\pi_{+})^{-1}(1+\ddot{f}\pi_{+}/\dot{f})^{-2}\frac{\delta^4(x_1-x_2)}{a^3}\nonumber\\
&=&\nu_{\calo}^{-1}\left(1-(\epsilon_{\nu_{\calo}}+2\epsilon_f)H\pi_{+}\right)\frac{\delta^4(x_1-x_2)}{a^3}.
\end{eqnarray}

This analysis provides a formal support to the manipulations behind Eqs. (\ref{noisemix}) and (\ref{noisepi}).

\end{document}